\documentstyle[emulateapj]{article}
\def\gsim{\;\rlap{\lower 2.5pt
 \hbox{$\sim$}}\raise 1.5pt\hbox{$>$}\;}
\def\lsim{\;\rlap{\lower 2.5pt
   \hbox{$\sim$}}\raise 1.5pt\hbox{$<$}\;}
\def\HH{{\rm H$_2$}}
\def\etal{{\sl et al.}\,~}
\def\eV{{\rm \,eV}}
\journalid{}{}
\articleid{}{}

\makeatother

\begin{document}
\begin{flushright}
{\footnotesize
FERMILAB-Pub-99/041-A}
\end{flushright}
\nopagebreak
\vspace{-\baselineskip}
\title{The Radiative Feedback of the First Cosmological Objects}

\author{Zolt\'an Haiman$^1$, Tom Abel$^{2,3}$
\& Martin J. Rees$^4$ \\ \ \\
{\sl \footnotesize $^1$ NASA/Fermilab Astrophysics Group, Fermi
National Accelerator Laboratory, Batavia, IL 60510, USA \\
$^2$ LCA, NCSA,
University of Illinois at Urbana/Champaign, 405 N. Mathews Ave., 
Urbana, IL 61801, US \\
$^3$ Max-Planck-Institut f\"ur Astrophysik,
Karl-Schwarzschild-Stra{\ss}e 1, 85748 Garching, Germany \\
$^4$ Institute of Astronomy, Madingley Road, Cambridge, CB3 0HA, U.K.}}

\begin{abstract}
In hierarchical models of structure formation, an early cosmic UV background
(UVB) is produced by the small ($T_{\rm vir}\lsim 10^4$K) halos that collapse
before reionization.  The UVB at energies below 13.6eV suppresses the formation
of stars or black holes inside small halos, by photo--dissociating their only
cooling agent, molecular ${\rm H_2}$.  We self--consistently compute the
buildup of the early UVB in Press--Schechter models, coupled with ${\rm H_2}$
photo--dissociation both in the intergalactic medium (IGM), and inside
virialized halos.  We find that the intergalactic ${\rm H_2}$ has a negligible
effect on the UVB, both because its initial optical depth is small ($\lsim
0.1$), and because it is photo--dissociated at an early stage.  If the UV
sources in the first collapsed halos are stars, then their UV flux suppresses
further star--formation inside small halos.  This results in a pause in the
buildup of the UVB, and reionization is delayed until larger halos ($T_{\rm
vir}\gsim 10^4$K) collapse.  If the small halos host mini--quasars with hard
spectra extending to $\sim$1keV, then their X--rays balance the effects of the
UVB, the negative feedback does not occur, and reionization can be caused by
the small halos.
\end{abstract}


\section{Introduction}
\thispagestyle{empty}
Recent progress in high redshift observations has made it possible to estimate
the global star--formation history of the universe to redshifts as high as
$z\sim 5$.  Although the evolution of the star--formation rate (SFR) is
relatively well--established at lower redshifts ($z\lsim 2$), there is still
substantial uncertainty about the $2\lsim z \lsim 5$ era.  A study of Lyman
break galaxies in the Hubble Deep Field shows a decline of the SFR towards high
$z$, indicating a peak in the SFR around $z\sim 2$ (Madau et al. 1996).  On the
other hand, a recent census of $z\gsim 3$ galaxies, detected in Ly$\alpha$
emission in a $\sim 200$ times wider field, suggests that the SFR is either
constant, or increasing from $z\sim2$ to $z\sim5$ (Steidel 1999).  At present,
these uncertainties render extrapolations of the observed SFR to still higher
($z\gsim5$) redshifts unreliable.

We do know, however, that some stellar activity must have taken place at $z>5$.
The presence of heavy elements observed in the high redshift Ly$\alpha$ forest
(Tytler, Cowie), as well as the reionization of the intergalactic medium (IGM),
require sources of metals and ionizing photons in addition to the currently
known population of galaxies and quasars.  Such ultra--high redshift objects
are indeed expected in the current best--fit cosmological cold dark matter
(CDM) models, which predict that the first virialized halos appear at redshifts
as high as $z\sim 30$.  It is natural to identify these small halos as the
sites where the first stars or massive black holes are born, forming the first
generation of "mini--galaxies" or "mini--quasars".

The first collapsed halos have masses near the cosmological Jeans mass,
$M\sim10^{4-5}~{\rm M_\odot}$, implying a virial temperature of a few hundred
K.  In order to form either stars or black holes, the gas in these halos must
be able to cool efficiently.  At this low temperature, the only mechanism in
the chemically simple primordial gas that satisfies this requirement is
collisional excitations of ${\rm H_2}$ molecules.  At still higher temperatures
($T\gsim10^4$K) and masses ($M\sim 10^8~{\rm M_\odot}$), efficient cooling is
enabled by atomic hydrogen. The presence or absence of ${\rm H_2}$ molecules in
the first collapsed halos therefore determines whether these small halos can
form any stars or quasars, before larger halos dominate the collapsed baryonic
fraction.  As a consequence, the ${\rm H_2}$ abundance determines when the
"dark age" ends: if there is sufficient ${\rm H_2}$ inside the first small
halos, the dark age ends earlier; if ${\rm H_2}$ is absent, it ends at the
later redshift when the larger scales collapse.

The fraction of ${\rm H_2}$ molecules in the post--recombination IGM is $x_{\rm
H_2}\equiv n_{\rm H2}/n_{\rm H}\sim 10^{-6}$.  In the absence of a background
radiation field, this fraction rises to $\sim10^{-4}$ at the dense central
regions of collapsed halos (Haiman et al. 1996; Tegmark et al. 1997), high
enough to satisfy the cooling criterion.  However, ${\rm H_2}$ molecules are
fragile, and are photo-dissociated efficiently by a low--intensity UV
radiation.  In a previous paper (Haiman et al. 1997, hereafter HRL97), we
showed that the flux necessary for ${\rm H_2}$ photo--dissociation is several
orders of magnitude smaller than the flux needed to reionize the universe.
Based on this result, we conjectured that the ${\rm H_2}$ abundance in the
first collapsed halos was suppressed soon after a small number of these halos
formed stars or turned into mini--quasars.

This negative feedback would imply that the bulk of the first mini--galaxies or
mini--quasars would reside in dark halos with masses at least $10^{7-8}~{\rm
M_\odot}$, and that the "dark age" would end only at the typical collapse
redshift of these halos.  The first condensations at the $\sim 1000$ times
smaller Jeans mass would then remain neutral, gravitationally confined gas
clumps, until they merge into larger systems, or are photo--evaporated by the
UVB established after reionization (Barkana \& Loeb 1998).  A small fraction of
these clumps is expected to survive mergers and photo--evaporation until $z\sim
3$; based on their column density and abundance, Abel \& Mo (1998) have
identified this survived fraction with the Lyman limit systems observed near
$z\sim 3$.  A direct observational study of the first collapsed halos will be
feasible in the future with the Next Generation Space Telescope (NGST): with
its $\sim 1$nJy imaging sensitivity, NGST will be able to detect a $\sim
10^6~{\rm M_\odot}$ halo with an average star formation efficiency at $z\sim
10$ (Smith \& Koratkar 1998).

In this paper, we study the conjecture of HRL97 in a cosmological setting, by
quantifying the radiative feedback of the first collapsed halos on their ${\rm
H_2}$ abundance, as the early UVB is established in Press--Schechter models.
The two main questions we address are: {\it (1) How does the ${\rm H_2}$
feedback effect the typical masses and collapse redshifts of the first halos};
and {\it (2) Could the first collapsed dark halos, with masses near the Jeans
mass, contribute to the early UVB, and to the reionization of the universe?}

In HRL97, we concluded that the negative feedback suppresses the
fragmentation and collapse of small halos.  In the present paper, we
confirm this basic conclusion, although we find that it depends
sensitively on the spectrum of the early sources.  Our calculations
improve HRL97 in several ways.  We solve the coupled evolution of the
formation of new halos (using the Press--Schechter formalism), the
build--up of the early UVB, and the evolution of the ${\rm H_2}$
abundance in the IGM and inside each collapsed halo; we explore the
sensitivity of our conclusions on the spectral shape, specifically the
X--ray to UV flux ratio; we replace our assumption of homogeneous
slabs by the centrally condensed profiles expected for virialized
halos; and we employ a more realistic star--formation criterion.

This paper is organized as follows.  In \S~2, we quantify the spectrum
of the early UVB after processing through the intergalactic H and
${\rm H_2}$.  In \S~3, we compute the ${\rm H_2}$ abundance in
virialized halos under this UVB, and define our criterion for
star--formation.  In \S~4, we compute the evolution of the UVB coupled
with the ${\rm H_2}$--feedback, and present these results in \S~5.  In
\S~6, we discuss how these results are modified in the presence of
early mini--quasars with spectra extending to the X--rays.  Finally,
\S~7 summarizes the main conclusions and implications of this work.

Throughout this paper, we adopt a background $\Lambda$CDM cosmology with a
tilted power spectrum,
$(\Omega_0,\Omega_\Lambda, \Omega_{\rm  b}, h, \sigma_{8h^{-1}}, n)=
(0.35, 0.65, 0.04, 0.65, 0.87, 0.96)$;
however, our results are relevant in any model where the cosmic structure is
built hierarchically from the ``bottom up''.  For convenience, we also adopt
the following terminology: a "small halo" is defined to be a halo with virial
temperature $T\lsim10^4$K, or equivalently, with mass $M\lsim3\times 10^7~{\rm
M_\odot}[(1+z)/11]^{-3/2}$.  Conversely, a "large halo" is defined as a halo
with virial temperature $T\gsim10^4$K, or equivalently, with mass
$M\gsim3\times 10^7~{\rm M_\odot} [(1+z)/11]^{-3/2}$.

\section{Radiative Transfer of the Soft UV Background}

In hierarchical models of structure formation, small objects appear at high
redshift, and larger objects form later, with the total fraction of collapsed
baryons monotonically increasing with time.  The radiation output associated
with the collapsed halos gradually builds up a cosmic UV background.  Of
interest here is the background in the 11.18--13.6 eV range, enclosing the
Lyman and Werner (LW) bands of molecular ${\rm H_2}$.  Photons in this range do
not ionize neutral H atoms, and can travel to a large fraction of the Hubble
length across the IGM, and photo--dissociate molecular ${\rm H_2}$ both in the
IGM and inside other collapsed halos.  Since the presence of ${\rm H_2}$ is a
necessary condition for the formation of stars or a central black hole, this
leads to a feedback that controls the fraction of collapsed baryons turned into
luminous sources, and the evolution of the UVB.  Before addressing the feedback
itself, in this section we quantify the spectral shape of the UVB in the
relevant frequency range.  This is determined by the absorption of the flux by
intergalactic H and ${\rm H_2}$.

\subsection{"Sawtooth" Modulation From Neutral H}

\begin{figure*}[t]
\plotone{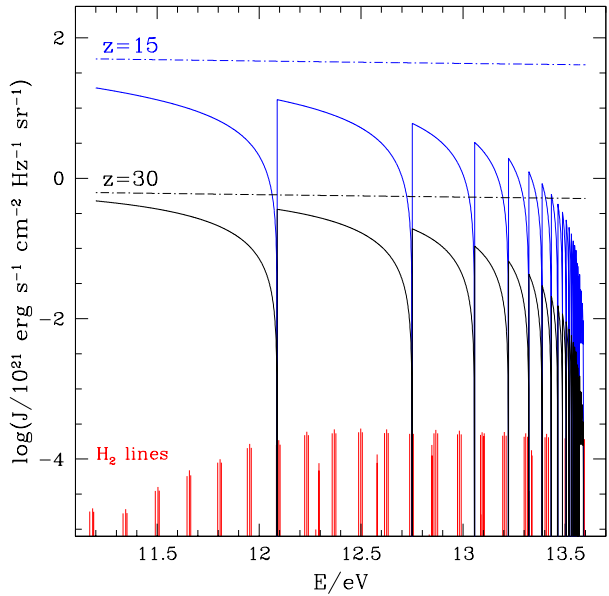}
\caption[Effect of intergalactic H2 on UVB] {\label{fig:sawtooth} 
\footnotesize The spectrum
of the UV background at redshifts $z=30$ and $z=15$ in the cosmological model
of Figure~\ref{fig:Res1}.  The solid lines show the spectrum after absorption
by neutral H in the high--redshift IGM; the dashed lines show the spectrum
without this absorption.  For reference, at the bottom of the figure we show
the location of the 76 LW lines of ${\rm H_2}$.  The length of each line
indicates its relative contribution to ${\rm H_2}$ dissociations (see the
Appendix and Figure~\ref{fig:LW} for the numerical values of the dissociation
fractions and oscillator strengths.}
\end{figure*}

Consider an observer at redshift $z_{\rm obs}$, measuring the UVB at the local
frequency $\nu_{\rm obs}$, where $11.18{\rm eV} < h\nu_{\rm obs} < 13.6{\rm
eV}$.  The photons collected at the frequency $\nu_{\rm obs}$ are arriving from
sources located at $z_{\rm s}>z_{\rm obs}$, and have redshifted their original
frequency $\nu_{\rm s}>\nu_{\rm obs}$ to $\nu_{\rm obs}$.  If during its
travel, the photon's frequency equals the frequency of any atomic Lyman line,
it is absorbed by the neutral IGM, because the optical depth of the IGM in
these lines are exceedingly high, e.g. $\tau_{\rm Ly\alpha} \sim10^6$ at
$z\sim20$.  This statement applies to all Lyman lines with $2<n\lsim 150$.  The
Ly$\alpha$ absorption ($n=1$) has only a small effect on the background, since
most of the absorbed photons are simply re-emitted at the same frequency,
except those converted to lower frequency photons by two--photon decays.  In
any case, the Ly$\alpha$ resonance is at 10.2 eV, outside the frequency range
of interest here. The very high Lyman lines ($n\gsim 150$) have small enough
cross--sections so that the total optical depth of the IGM at these frequencies
drops below unity. However, these resonances have a negligible contribution to
the total absorption in the 11.8-13.6eV range.  Here we make the simple
assumption that a "dark screen" blocks the view of all sources at redshifts
above $z_{\rm max}$. The maximum redshift is given by
\begin{equation}
\frac{1+z_{\rm max}}{1+z_{\rm obs}}=
\frac{\nu_{\rm i}}{\nu_{\rm obs}},
\label{eq:screen}
\end{equation}
where $\nu_{\rm i}$ is the frequency of the Lyman line closest from
above to the observed frequency, $\nu_{\rm obs}$.

The processing of the early UVB by the neutral IGM leads to a
characteristic spectral shape, resembling a "sawtooth", since the
closer $\nu_{\rm obs}$ is to a resonance, the nearer the screen is to
the observer.  The resulting spectrum is shown in a specific model
(described below) in Figure~\ref{fig:sawtooth}.  The redshift
distribution of sources is important for the magnitude of this
``sawtoothing'' effect.  If a large fraction of the sources
contributing to the UVB are located near the observer, then the effect
is smaller; if the sources are spread out over a wider range of
redshifts, then the effect is larger.  In the hierarchical models, new
halos form exponentially rapidly at high redshifts.  At any given
time, most of the sources are expected to have formed recently, i.e.
at redshifts only slightly above that of the observer.  Since the
formation rate of new halos decreases with redshifts, the sawtoothing
will be increasingly pronounced at later times.  This is illustrated
in Figure~\ref{fig:sawtooth}: the flux is suppressed by a larger
average factor at $z=15$ than it is at $z=30$.

\subsection{Modulation From ${\rm\bf\protect H_2}$ Molecules}

\begin{figure*}[t]
\plotone{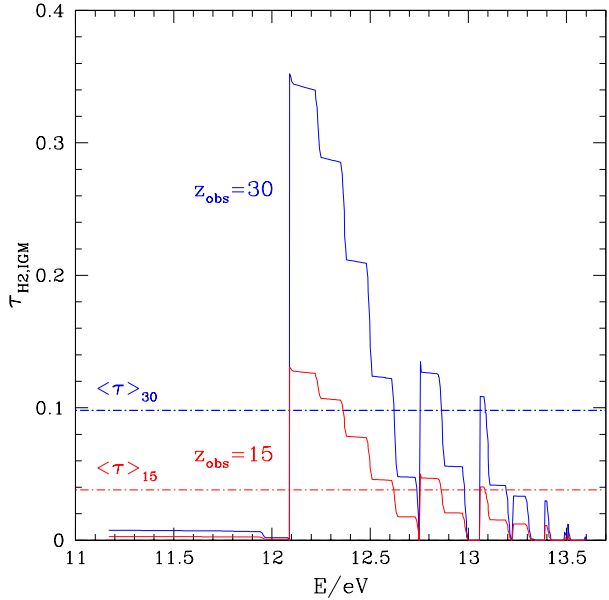}
\caption[Effect of intergalactic H2 on UVB] {\label{fig:igmh2}
  \footnotesize The total ${\rm H_2}$ optical depth of the IGM for an
  observer at redshift $z=30$ and $z=15$.  We assume a constant fixed
  $x_{\rm H_2}=2\times10^{-6}$, and sum the optical depth of the
  Lyman-Werner lines of ${\rm H_2}$.  At each observed frequency, we
  only include sources between $z_{\rm obs}$ and the nearest atomic
  Lyman resonance.  The dashed lines show the effective optical depths
  averaged in the 11.18--13.6eV range.}
\end{figure*}

In addition to absorption by neutral H, the UVB may also be modulated
by the intergalactic ${\rm H_2}$.  The ${\rm H_2}$ in the IGM removes
photons from the UVB when these molecules are photo--dissociated, and
decreases the rate of photo--dissociations inside collapsed halos.  It
is important, therefore, to assess the magnitude of the modulation
from the intergalactic ${\rm H_2}$.

When a LW photon is absorbed by an ${\rm H_2}$ molecule, there is on average a
$\sim15$\% chance that it causes a photo--dissociation of the molecule.  This
is the average probability that the molecule decays from its excited state to a
continuum of two distinct H atoms (Solomon, cf. Field et al. 1966; Stecher \&
Williams 1967).  We will assume here that in the other $\sim85\%$ of
absorptions, the LW photons is re-emitted at the same frequency, and there is
no net effect.  This is justified because the quadrupole transitions among the
excited electronic states are much slower than decay back to the electronic
ground state, and also because the small dipole moment that preferred the upper
state in the absorption will preferentially populate the initial lower state in
the decay (Herzberg 1950).

If the ${\rm H_2}$ fraction in the IGM is $x_{\rm H_2}=n_{\rm H_2}/n_{\rm H}$,
then the total optical depth of the IGM at redshift $z$ and frequency $\nu$
to ${\rm H_2}$--dissociations is given by
\begin{equation}
\tau_{\rm H_2}(\nu,z)=x_{\rm H_2} \frac{\pi e^2}{m_{\rm e} c} 
\sum_{i=1}^{76} f_{\rm diss, i} f_{\rm osc, i} 
\int_z dz^\prime c \frac{dt}{dz^\prime} n_{\rm H}(z^\prime) 
\phi(\nu_{z^\prime}, \nu_{\rm i}),
\label{eq:tauH2a}
\end{equation}
where the sum is over the 76 LW lines of ${\rm H_2}$; $n_{H}=8.5 \times 10^{-6}
\Omega_{\rm b} h^2 (1+z)^3$ is the neutral H number density in the IGM; $f_{\rm
diss, i}$ is the probability that absorption into the ${\rm i^{th}}$ LW line is
followed by a dissociative decay; $f_{\rm osc, i}$ is the oscillator strength
of the ${\rm i^{th}}$ LW line; $\nu_{z^\prime}\equiv\nu(1+z^\prime)/(1+z)$;
$cdt/dz$ is the line element in our $\Lambda$CDM cosmology; and $\phi(\nu,
\nu_{\rm i})$ is the absorption line profile centered around the resonance
frequency $\nu_{\rm i}$.  We included the equilibrium ortho to para ratio of
3:1 of \HH~implicitly in equation~(\ref{eq:tauH2a}) by appropriately
multiplying the oscillator strengths with 0.75 or 0.25
respectively. Furthermore, we assume that the line shapes follow a Voigt
profile with thermal broadening by the IGM temperature (taken to be
$T=0.0135(1+z)^2$\,K, although the total optical depth, and therefore our
results, are insensitive to this choice). More information on the details of
\HH~pre--dissociation are given in the Appendix.

Equation~\ref{eq:tauH2a} gives a formal sum total optical depth of the 76 LW
resonances, each of which is at a different frequency.  From the point of view
of a photon traveling across the IGM, these resonances would be encountered at
different redshifts. During its lifetime, a UV photon can therefore be
subjected only to a subset of the LW resonances: as we argued in \S~2.1 above,
the maximum redshift interval a UV photon can travel before it is absorbed by a
neutral H atom corresponds to the redshift between two neutral H resonances.  A
simple calculation of the optical depth using equation~\ref{eq:tauH2a} would
result in an overestimate by a factor of a few (depending on $z$ and $\nu$) of
the true value relevant to the observed modulation of the UVB.  To impose the
fact that the UVB is {\it simultaneously} modulated both by the intergalactic H
and ${\rm H_2}$, one needs to change the summation in Equation~\ref{eq:tauH2a}
from $i\in$(1,76) to $i\in$($i_{\rm min}$,$i_{\rm max}$), where $i_{\rm min}$
is the LW resonance closest in frequency from above to $\nu$, and $i_{\rm max}$
is the LW resonance farthest above $\nu$ before encountering an atomic Lyman
line (see Figure~\ref{fig:sawtooth} and Figure~\ref{fig:LW} for visualizations
of the relative location of the H and ${\rm H_2}$ lines).
 
In Figure~\ref{fig:igmh2}, we show the resulting optical depth as a function of
the observed frequency, for two different redshifts, $z=15$ and $z=30$.  As the
figure shows, the optical depth is largest at $\sim$12.1eV.  This is because
the largest frequency gap between the atomic lines occurs between Ly$\beta$ and
Ly$\gamma$, enclosing the largest subset of the ${\rm H_2}$ LW lines with
significant contributions to ${\rm H_2}$ dissociations (see
Figures~\ref{fig:sawtooth} and \ref{fig:LW}).  Figure~ \ref{fig:igmh2} also
shows that the maximum optical depth is 0.35 and 0.13 for $z=30$ and $z=15$,
respectively (assuming $x_{\rm H_2}=2\times10^{-6}$).  Interestingly, the
characteristic features in Figure~\ref{fig:igmh2} depend only on the relative
frequencies of atomic vs. molecular resonant absorption lines (shown in
Figure~\ref{fig:LW}), i.e. only on atomic and molecular physics.

The total optical depths are near, but somewhat below unity.  This result has
several consequences.  It is useful here to draw an analogy with the ionization
of neutral H.  The total optical depth of the IGM to H photo--ionizations (just
above 13.6eV) is $\tau_{\rm H}\sim100 [(1+z)/21]^3$, which justifies the simple
picture of isolated HII regions ("Str\"omgren spheres"), each bounded by a
sharp ionization front, and continuously expanding around each ionizing source.
The situation with ${\rm H_2}$ is different in two ways: the optical depth is
less than unity, and is distributed among several LW lines, each of which is
individually optically thin, and causes dissociations at a different redshift
relative to each source.  As a consequence, no sharp ${\rm H_2}$ dissociation
``front'' exists.

\begin{figure*}[t]
\plotone{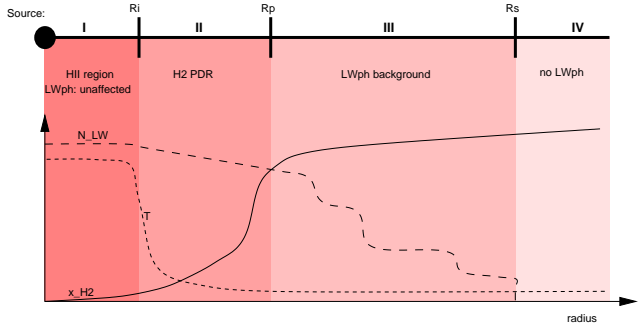}
\caption[Radial structure of Cosmological PDR.]  {\label{fig:pdr}
  \footnotesize Schematic sketch of the radial structure of a
  cosmological photo--dissociation region. Region I is the usual
  Str\"omgren sphere, bounded by the \ion{H}{1} ionization front at
  $R_{\rm i}$. Inside this region, the temperature is high
  ($\sim10^4$K) hydrogen is ionized, and the \HH~abundance tends to
  zero.  Region II extends from $R_{\rm i}$ to the radius $R_{\rm p}$
  where the \HH~photodissociation time becomes longer than the Hubble
  time.  Here the temperature is that of the IGM, the \HH~fractions
  are still low, but slowly rise to the background value of $\sim
  2\times 10^{-6}$. Region III extends from $R_{\rm p}$ to the radius
  $R_{\rm s}$, where Ly$\gamma$ shifts to Lyman$\beta$.  In this
  region, the source contributes to the soft UV background, but its
  flux is diminished by absorption into the atomic Lyman series. In
  region IV beyond $R_{\rm s}$, all photons in the Lyman Werner bands
  are absorbed by an atomic Lyman line. Hence, at such large scales,
  an individual source does not contribute to the UVB and
  intergalactic \HH~dissociation. See \S~2.2 for a more detailed
  discussion.}
\end{figure*}

A schematic sketch of four distinct regions in the radial structure of an ${\rm
H_2}$ photo--dissociation region is shown in Figure~\ref{fig:pdr}.  Region I
extends to the usual \ion{H}{1} ionization front or Str\"omgren radius. Inside
this region, the temperature is high ($\sim10^4$K), hydrogen is ionized, \HH~is
photo--dissociated, and the \HH~abundance tends to zero.  Region II is defined
to extend from $R_{\rm i}$ to the radius $R_{\rm p} \sim 2.5 \, h^{-1/2}
(1+z)_{30}^{-3/4} N_{LW,47}^{1/2} \;\; {\rm kpc}$, where the
\HH~photodissociation time becomes longer than the Hubble time. Here
$N_{LW,47}=N_{LW}/10^{47} {\rm s}^{-1}$ is the photon number flux in the Lyman
Werner bands per second. In this region the temperature is that of the IGM, and
the \HH~fractions are still low, but slowly rise to its background value of
$\sim 2\times 10^{-6}$. Since the optical depth in the Lyman Werner lines are
very small, the dissociation front is significantly diluted.  Region III is
defined to extend from $R_{\rm p}$ to the radius $R_{\rm s}\approx 2 h^{-1}
[(1+z)/31]^{-1.5}~{\rm Mpc}$ where Lyman $\gamma$ shifts to Lyman$\beta$.  In
this region, the source contributes to the soft UV background, but its flux is
diminished by absorption into the atomic Lyman series. In region IV beyond
$R_{\rm s}$, all photons in the Lyman Werner bands are absorbed by an atomic
Lyman line. Hence, at such large scales, an individual source does not
contribute to the UVB and intergalactic
\HH~dissociation.

Although the maximum optical depths from the intergalactic ${\rm H_2}$ are
relatively small, the UVB flux can be suppressed by as much as
$1-\exp(-0.35)\sim 50$\%.  However, we are interested in the suppression of the
${\rm H_2}$ photo--dissociation rate when the UVB falls on a collapsed halo.
This suppression is smaller than 50\%, since the photo--dissociation rate is
given by a weighted average of the observed flux in the 11.18--13.6eV range.
We therefore define an ``effective'' optical depth $<\tau>_z$, that directly
characterizes the suppression of the ${\rm H_2}$ photo--dissociation rate, as
follows:
\begin{equation}
\begin{array}{l}
k_{\rm diss,0}(z)  =  4\pi x_{\rm H_2} n_{\rm H}(z) \frac{\pi
  e^2}{m_{\rm e} c}  \times \\
\mbox{  } \sum_{\rm i=i_{min}}^{\rm i_{max}} f_{\rm diss, i} f_{\rm osc, i} 
\int_\nu d\nu^\prime \phi(\nu^\prime, \nu_{\rm i}) 
\frac{J_{\nu^\prime}(z)}{h\nu^\prime}, 
\end{array}\label{eq:dissoc0}
\end{equation}
\begin{equation}
\begin{array}{l}
k_{\rm diss}(z)  =4\pi x_{\rm H_2} n_{\rm H}(z) \frac{\pi e^2}{m_{\rm
  e} c}  \times \\
\sum_{\rm i=i_{min}}^{\rm i_{max}} f_{\rm diss, i} f_{\rm osc, i} 
\int_\nu d\nu^\prime \phi(\nu^\prime, \nu_{\rm i}) 
\frac{J_{\nu^\prime}(z)}{h\nu^\prime} \exp[-\tau_{\rm
  H_2}(\nu^\prime,z)], \label{eq:dissoc}
\end{array}
\end{equation}
\begin{equation}
<\tau>_z\equiv -\log[k_{\rm diss}(z)/k_{\rm diss,0}(z)].
\label{eq:tauave}
\end{equation}
Here $J_{\nu}(z)$ is the UVB flux before processing by the intergalactic ${\rm
H_2}$, and $\tau_{\rm H_2}(\nu,z)$ is the optical depth at this frequency shown
by the solid lines in Figure~\ref{fig:igmh2}.  When the sources are assumed to
be distributed according to the Press--Schechter model (see below), the maximum
reduction in the ${\rm H_2}$ photo--dissociation rate is 10\% at $z=30$ and 4\%
at $z=15$ (see the dashed lines in Figure~\ref{fig:igmh2}).  Note, however,
that these numbers assume a constant $x_{\rm H_2}=2\times10^{-6}$. In reality
(as will be shown below), the intergalactic ${\rm H_2}$ fraction drops sharply
soon after the appearance of the first sources, before the UVB builds up to the
level of causing any ${\rm H_2}$ feedback inside the collapsed clumps.  This is
because the ${\rm H_2}$ dissociation time in the IGM, $1/k_{\rm diss}$ becomes
shorter than the Hubble time at $J\sim 10^{-3}$, while the feedback on the
dense clumps typically turns on only later, when the UVB reaches $J\sim
10^{-2}$.  

In summary, we have shown that the intergalactic ${\rm H_2}$ might cause a
maximum of $\sim50\%$ suppression of the UVB near 12 eV, but the overall effect
on the ${\rm H_2}$ photo--dissociation rate is small: less than 10\%, assuming
a constant intergalactic ${\rm H_2}$ fraction of $2\times10^{-6}$.  We will
show below that the intergalactic ${\rm H_2}$ fraction is substantially
reduced, and therefore this small modulation disappears, by the redshifts of
interest -- when the UVB builds up to the levels to photo--dissociate ${\rm
H_2}$ molecules in dense collapsed clumps.

\section{${\rm\bf\protect H_2}$-Cooling in the Presence of the UV Background}

In this section, we consider the behaviour of individual, centrally condensed
gas clouds, when illuminated by an external UVB.  Here, we assume the UVB to
have a fixed amplitude, and postpone the treatment of the coupled evolution of
the UVB, and the ${\rm H_2}$ abundance and star formation to \S~4 below.

\subsection{The ${\rm\bf\protect H_2}$ Abundance in Collapsed Clouds}

The radiative efficiency of a gas cloud by collisional excitations of ${\rm
H_2}$ molecules depends on the gas temperature, density, and ${\rm H_2}$
abundance.  Although the typical average overdensity of a virialized spherical
perturbation is $\delta\sim 100-200$, in a centrally condensed cloud, the
central density can reach a much higher value.  This is important in the
present context, because the formation of ${\rm H_2}$ molecules can be enhanced
by the accelerated chemistry inside the central, dense regions.  In addition,
the total integrated column density of a centrally condensed cloud is higher
than that of a homogeneous sphere with the same mass and density. Central
condensation therefore tends to make the clouds more self--shielding against
the external UVB, and help to preserve the internal ${\rm H_2}$ molecules.

\begin{figure*}[t]
\plotone{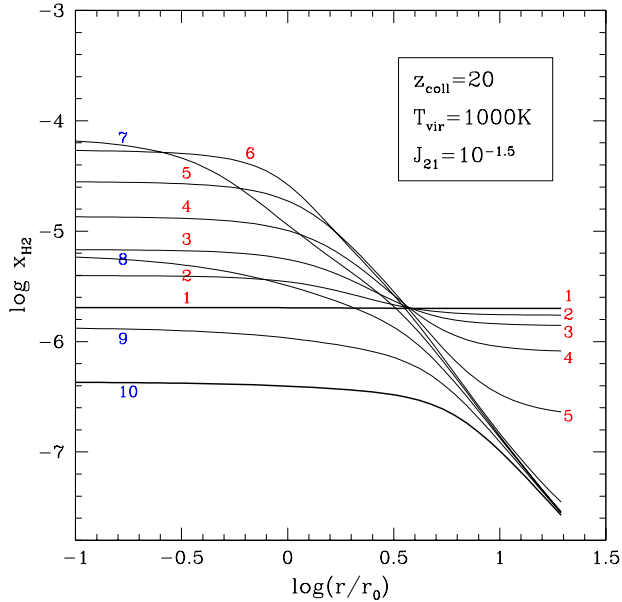}
\caption[The time-evolution of the H2-profile] {\label{fig:prof}
  \footnotesize The time--evolution of the ${\rm H_2}$ profile in a
  truncated isothermal sphere in the presence of an external flux.
  The example shown is for a truncated singular isothermal sphere with
  $T_{\rm vir}=10^3$K, $z_{\rm coll}=20$, and $J_{21}=10^{-1.5}$.
  Increasing numbers correspond to equal time--steps in logarithmic
  time, upto a Hubble time of $\sim 5\times10^{15}$ sec. At the
  outskirts of the cloud the \HH~fraction decreases with time. In the
  core molecules are formed initially via $H^-\ + \ H \ \rightarrow \ 
  H_2\ + \ e^-$ at a rate exceeding the photo--dissociation rate. At
  the point where the electron fraction drops below $10^{-5}$
  photo--dissociation becomes dominant and \HH~is destroyed in the
  already self--shielding core. }
\end{figure*}

To characterize the central condensation, here we adopt the density profiles of
truncated isothermal spheres (TIS, Shapiro et al. 1999).  These profiles are
especially convenient, since they have a one--to--one correspondence with the
final virialized state of the spherical top--hats used in the Press--Schechter
formalism.  The spheres are isothermal at the virial temperature, their average
overdensity is $\sim130$, and their central overdensity in the core reaches
$\sim10^4$. The TIS profiles are similar to the spherically averaged universal
profiles found in numerical simulation of clusters (NFW, Navarro et al. 1998).
The main difference from NFW is that the TIS has a central core (the existence
of cores are indicated by recent simulations of galaxy groups, e.g. by Kravtsov
et al. 1998).

We assume that the TIS is composed of the nine species ${\rm H}$, ${\rm H^-}$,
${\rm H^+}$, ${\rm He}$, ${\rm He^+}$, ${\rm He^{++}}$, ${\rm H_2}$, ${\rm
H_2^+}$, and ${\rm e^-}$, with the initial abundances given by their
post--recombination values (Anninos and Norman 1996).  Our results are
insensitive to the precise initial conditions, such as the electron fraction
(${\rm x_e\approx 10^{-4}}$) and molecule fraction ($x_{\rm H_2}\approx2\times
10^{-6}$). Although these abundances represent the chemical composition of the
IGM, and we start with a cloud that is already collapsed and virialized, this
should not effect our final results.  In 1--D spherical simulations (Haiman,
Thoul \& Loeb 1996), we have found that the abundances stay near their initial
IGM values until the formation of the virialization shock; soon after
virialization the abundances loose memory of their initial values.

We solve the subsequent time evolution of the chemical
abundances, as a function of radius throughout the
cloud, assuming that the cloud is illuminated by the
external UVB.   The UVB flux is assumed to be a power
law below 13.6eV,
\begin{equation}
J_{\nu}=J_{21}\left(\frac{\nu}{\nu_{\rm H}}\right)^{-1}~\times
{\rm sawtooth~}~~~~h\nu<13.6
{\rm eV},  
\label{eq:flux}
\end{equation}
and to be zero above 13.6eV, since the flux from stellar sources would
suppressed by several orders of magnitude by the neutral IGM.  The sawtooth
modulation by the intergalactic H is described above, and the constant
amplitude $J_{21}$ is in units of $10^{-21}~~{\rm
erg~s^{-1}~cm^{-2}~Hz^{-1}~sr^{-1}}$.  Our results are insensitive to the shape
of the UVB, since we only utilize the flux between 11.18 and 13.6 eV. The
possibility of a non--stellar flux, extending to higher energies will be
discussed in section \S~6 below. The chemical reaction rates and
cross--sections are taken from Haiman, Thoul \& Loeb (1996).  We have solved
the radiative transfer across the cloud, assuming spherical symmetry, and
included the self--shielding due to both H and ${\rm H_2}$.  The initial
conditions are specified by the virial temperature $T_{\rm vir}$, and collapse
redshift $z_{\rm coll}$ of the TIS, and the amplitude of the UVB, $J_{21}$.

\begin{figure*}[t]
\plotone{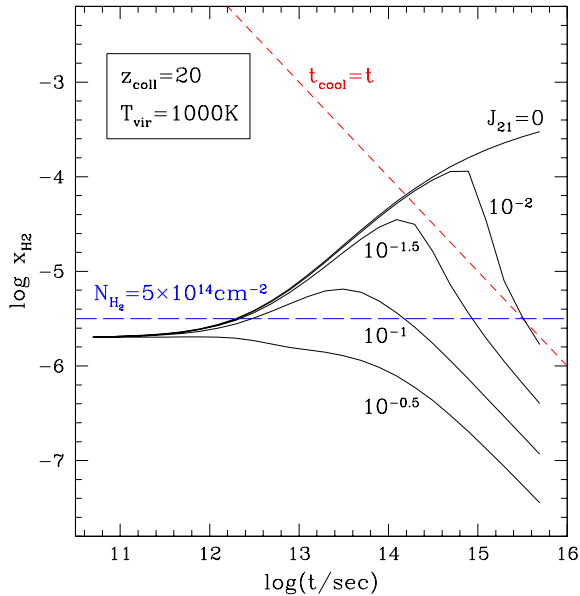}
\caption[Evolution of H2-fraction in the core.]  {\label{fig:evol}
  \footnotesize Evolution of the ${\rm H_2}$ abundance in the core of
  the singular isothermal sphere shown in Figure~\ref{fig:prof}; in
  the presence of various UV background fluxes (solid lines).  Also
  shown is the ${\rm H_2}$ abundance at which the cooling time equals
  the time elapsed since the collapse of the cloud (short dashed
  line); and the ${\rm H_2}$ abundance at which the LW lines would
  become self--shielding (at an integrated ${\rm H_2}$ column density
  of $5\times 10^{14}~{\rm cm^{-2}}$, long dashed line).  For
  $J>10^{-1.5}$ the cloud becomes self--shielding, but still cannot
  cool fast enough.}
\end{figure*}

An illustrative example of the evolution of the ${\rm H_2}$ abundance is shown
in Figures~\ref{fig:prof} and~\ref{fig:evol}.  Figure~\ref{fig:prof} shows the
radial profile of $x_{\rm H_2}$ at 10 different time--steps, equally spaced in
logarithmic time, during a time--interval that corresponds to the Hubble time
at $z=20$, $t\approx 5\times 10^{15}$ sec.  The cloud is assumed to be a TIS
with $z_{\rm coll}=20$ and $T_{\rm vir}=10^3$K, and the amplitude of the UVB is
taken to be $J_{21}=10^{-1.5}$.  As Figure~\ref{fig:prof} shows, initially the
${\rm H_2}$ fraction rises in the core: this is because new ${\rm H_2}$
molecules form via the charge transfer reactions ${\rm H + e^- \rightarrow H^-
+ \gamma}$ and ${\rm H^- + H \rightarrow H_2 + e^-}$, utilizing the initial
residual free electrons.  By $t=10^{14}$ sec, $x_{\rm H_2}$ reaches a value
near $\sim10^{-4}$ (cf. the lines labeled 1--6).  At this point, however, the
free electron fraction has dropped by a factor of $\sim2$, and although the
core is self--shielding, photo--dissociation becomes faster than
\HH~formation. As a consequence $x_{\rm H_2}$ drops continuously, to a level
below the initial $\sim10^{-6}$ by $t=10^{15}$ sec (cf. the lines labeled
7--10), while $x_{\rm e}$ drops to $\sim2\times 10^{-6}$.

The evolution of $x_{\rm H_2}$ at the core radius $r=r_0$ is also shown by the
middle solid curves in Figure~\ref{fig:evol} for various $J_{21}$.  The shell
at $r=r_0$ is normally representative of the whole core, because the density,
as well as $x_{\rm H_2}$, is flat within $r<r_0$ (in some cases, however, the
${\rm H_2}$ fraction may still rise within the core, see \S~6 below).  For
comparison, in this Figure we also show the evolution of $x_{\rm H_2}$ in the
same shell under less intense ($J_{21}=0$ or $10^{-2}$) or stronger
($J_{21}=10^{-1}$ or $10^{-0.5}$) UV backgrounds.  In the absence of any UVB,
the ${\rm H_2}$ fraction rises continuously, and reaches $10^{-3}$ in a Hubble
time (top curve); when a flux is turned on and increased (bottom 4 solid
curves), the ${\rm H_2}$ fraction is continuously suppressed.

By performing similar calculations for clouds with virial temperatures and
collapse redshifts in the ranges $10^2 < T_{\rm vir} < 10^4$K and $5 < z_{\rm
coll}< 50$, we have found that the evolution shown in Figures~\ref{fig:prof}
and~\ref{fig:evol} remains qualitatively the same in all cases. However, the
level of flux required at which the ${\rm H_2}$ molecules are
photo--dissociation depends on the values of both $z_{\rm coll}$ and $T_{\rm
vir}$.

\subsection{The ${\rm\bf\protect H_2}$ Abundance and Criterion for
Star Formation}

In order for a cloud to fragment into stars, or continue collapsing to form a
central black hole, it is necessary for the gas to radiate efficiently.  A
rough estimate of the ``critical'' ${\rm H_2}$ fraction can be obtained by
requiring that the cooling time is shorter than either the dynamical time
(HRL97), or the Hubble time (Tegmark et al 1997).  Here we adopt a somewhat
more elaborate criterion for star formation, as follows.  For a given
combination of $z_{\rm coll}$, $T_{\rm vir}$, and $J_{21}$, the cloud is
assumed to be able to cool if, at any time during its evolution, the cooling
time in the core exceeds the ``present time'', defined as the time elapsed
since the formation of the cloud at $z_{\rm coll}$.  We assume this to be both
a necessary and a sufficient criterion.

This definition explicitly takes into account the behaviour of the ${\rm H_2}$
fraction shown in Figures~\ref{fig:prof} and~\ref{fig:evol}, especially the
fact that the ${\rm H_2}$ fraction in the core first rises and then declines
again.  Our criterion imposes the requirement that the cloud not spend more
than a cooling time at a given ${\rm H_2}$ fraction, before the ${\rm H_2}$
abundance is further reduced by photo--dissociation.  This definition avoids
star formation in a situation when the cooling time exceeds the dynamical time,
but only for a brief interval that would be too short for a significant
contraction to take place.  In Figure~\ref{fig:evol}, our criterion is
implemented by requiring that the solid line (the evolving ${\rm H_2}$
fraction) goes above the short--dashed line (the critical ${\rm H_2}$ fraction
at which $t_{\rm cool}$ equals the present time $t$).

\begin{figure*}[t]
\plotone{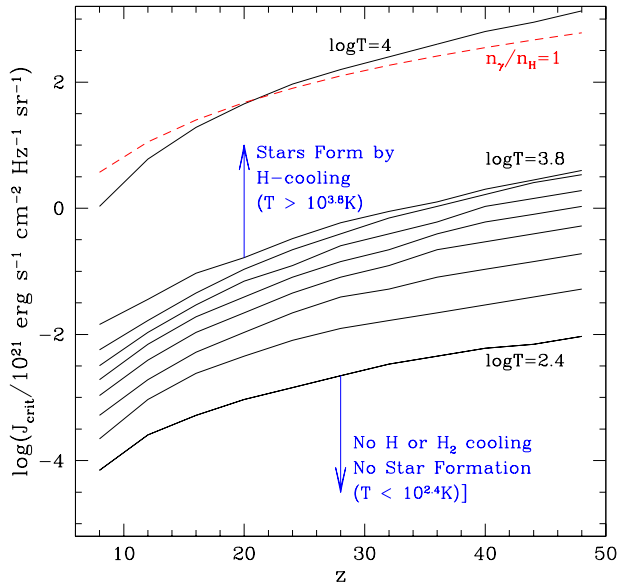}
\caption[Critical flux for star formation.]  {\label{fig:Jmin}
  \footnotesize The critical flux above which star formation in a halo
  with virial temperature $T_{\rm vir}$ and collapse redshift $z_{\rm
    coll}$ is prevented.  Halos smaller than $T_{\rm
    vir}\approx10^{2.4}$K cannot cool even in the absence of any flux.
  Halos larger than $T_{\rm vir}\approx10^{3.8}$K can cool via atomic
  H, irrespective of the ${\rm H_2}$ abundance.  For comparison, the
  dashed curve show the flux corresponding to one ionizing photon per
  H atom, the minimum flux necessary to reionize the universe.}
\end{figure*}

Interestingly, we find that the cloud may undergo a brief interval during which
it is self--shielding in the ${\rm H_2}$ LW bands, but then may become
optically thin again as the ${\rm H_2}$ abundance drops due to subsequent
photo--dissociation.  This is demonstrated in Figure~\ref{fig:evol}: the long
dashed line shows the value of $x_{\rm H_2}$ that would imply an integrated
${\rm H_2}$ column density of $N_{\rm H_2}=5\times10^{14}~{\rm cm^{-2}}$ across
the cloud, at which the LW lines become self--shielding.  As the figure shows,
if the flux is below $J_{21}\approx 10^{-0.7}$, the ${\rm H_2}$ fraction does
temporarily rise above this critical value; but for $J_{21}\gsim 10^{-1.5}$,
our criterion for star formation is still not satisfied.  This implies that
self--shielding by itself does not guarantee that the molecules are preserved
and enable efficient cooling. Requiring the cloud simply to self--shield would
therefore not be a good star formation criterion.  However, the decline in the
\HH~fraction is due to the decreasing electron abundance which lowers the
\HH~formation rate. Hence, if an additional process keeps the $e^-$ fraction
greater than $\sim 10^{-5}$, more \HH~could be formed. This possibility is
discussed in greater detail in \S~6.

\subsection{The Minimum Flux for Negative Feedback}

In addressing the effect of the feedback from the UVB, the main question is:
given a collapsed cloud with $z_{\rm coll}$ and $T_{\rm vir}$, what is the
critical value of the background flux, $J_{\rm crit}$ so that star--formation
in this collapsed cloud is suppressed?  In the example shown in
Figures~\ref{fig:prof} and~\ref{fig:evol}, $(z_{\rm coll},T_{\rm
vir})=(20,10^3{\rm K})$, $J_{\rm crit}$ can be read off directly from
Figure~\ref{fig:evol}, $J_{\rm crit}=10^{-1.5}$.  In order to be able to
investigate the evolution of the UVB in the Press--Schechter models, we have
repeated the same calculation for a grid of values for $5<z_{\rm coll}<50$, and
$2<\log(T_{\rm vir}/{\rm K})<4$.  For each pair of values, we have found the
critical flux for preventing star formation, $J_{\rm crit}$.

The resulting $J_{\rm crit}=J_{\rm crit}(z_{\rm coll},T_{\rm vir})$ are shown
in Figure~\ref{fig:Jmin}.  In general, the higher the density (or $z_{\rm
coll}$) or temperature, the higher the flux needs to be to prevent star
formation.  We have also found that for temperatures below $\log(T_{\rm
vir}/{\rm K})<2.4$, the value of the flux becomes irrelevant for star
formation, because the cooling function of ${\rm H_2}$ drops, and ${\rm H_2}$
cooling is inefficient at these low temperatures, even in the absence of any
flux.  Similarly, the value of the flux is irrelevant for $\log(T_{\rm
vir}/{\rm K})>3.8$, because at these high temperatures, cooling from
(collisional excitations of) neutral H always dominates over ${\rm H_2}$
cooling, i.e. stars can form irrespective of the ${\rm H_2}$ abundance.

The values of $J_{\rm crit}$ shown in Figure~\ref{fig:Jmin} are relatively low.
For reference, the dashed curve in this figure shows the value of $J_{\rm
crit}$ that corresponds to one ionizing photon per hydrogen atom, i.e. the
minimum flux necessary to reionize the universe.  {\it Since the values of the
flux necessary to prevent star--formation are 2--4 orders of magnitude below
this minimum reionizing flux, a negative feedback necessarily turns on prior to
reionization}.  If the star formation efficiency was a universal constant
inside all halos that can cool efficiently, independent of redshift and halo
size, then the negative feedback would turn on long before reionization -- when
the collapsed fraction of baryons $f_{\rm coll}$ is $\sim$4 orders of magnitude
smaller then it is at reionization -- and it would last until $f_{\rm coll}$
rises by a factor of $\sim100$.

Since new sources appear exponentially fast in the Press--Schechter formalism,
the relatively ``long'' interval in $f_{\rm coll}$ can translate into a
``short'' interval in redshift.  However, it is clear from the above that the
fundamental ``evolutionary parameter'' for the negative feedback, followed by
reionization, is $f_{\rm coll}$, rather than the redshift to which it
translates to.  Our results may therefore vary in different cosmologies when
stated in terms of redshift, but would be relatively robust to change in the
background cosmology, when stated in terms of $f_{\rm coll}$.  Finally, we
emphasize that the actual redshift (and $f_{\rm coll}$) interval between the
turn--on of the negative feedback and reionization would be shorter if the
star--formation efficiency in small halos is intrinsically much smaller than in
large halos (see discussion below).

\section{${\rm\bf\protect H_2}$--Feedback and the Buildup of the Cosmic UV Background}

In this section, we put together the results of the previous sections to compute
the redshift evolution of the UV background, the intergalactic ${\rm H_2}$
fraction, and the mass--scale of the ${\rm H_2}$ feedback.  The latter is more
conveniently expressed by defining a critical virial temperature, $T_{\rm
crit}(z)$, that corresponds to the minimum size of a cloud that satisfies our
star formation criterion under a fixed UV background $J_{21}$ and collapse
redshift $z_{\rm coll}$. The evolution of $T_{\rm crit}(z)$ and the UVB
amplitude, $J_{21}(z)$ are then coupled together, since $T_{\rm crit}(z)$ is
directly determined by $J_{21}(z)$, and conversely, $J_{21}(z)$ depends on the
$T_{\rm crit}(z)$ above which we allow halos to contribute to the UVB.  As
argued above, the evolution of the intergalactic ${\rm H_2}$ can be considered
separately, since it does not modulate the UVB at the level that would alter
the ${\rm H_2}$ photodissociation rate inside halos.

Our model is based on the Press--Schechter formalism.  We assume that the
mass--function of dark halos is given by the Press--Schechter formula, $N_{\rm
ps}(z,M)$.  The mass function of collapsed baryonic clouds is obtained from
$N_{\rm ps}(z,M)$ by the simple replacement $M\rightarrow \Omega_{\rm
b}/\Omega_{\rm tot} M$.  For the smallest masses, near the Jeans mass
($M\sim10^{4-5}{\rm M_\odot}$), we apply a correction to take into account the
non--zero gas pressure in these clouds.  This is achieved by replacing
$M\rightarrow f_{\rm b} \Omega_{\rm b}/\Omega_{\rm tot} M$, where $0\le f_{\rm
b} \le 1$ is the fraction of baryons that settle in the potential well of a
dark halo of mass $M$ by redshift $z$ (Haiman \& Loeb 1997).  We use the
mass--virial temperature relation of the TIS solutions.  Although this relation
was derived in a standard CDM cosmology, the scaling of $T_{\rm vir}$ with
redshift is changed only negligibly in our $\Lambda$CDM cosmology at the high
redshifts of interest.

To parameterize the uncertainties regarding the ionizing photon production rate
per collapsed baryon, we assume that each halo (provided it satisfies the star
formation criterion) has a constant star formation rate for $t_{\rm on}$ years,
during which it turns a total fraction $\epsilon$ of its baryonic mass into
stars with a Scalo IMF.  This parameterization could correspond to different
physical scenarios.  For instance, $\epsilon$ would remain the same if the star
formation rate in each halo was doubled, but the escape fraction of ionizing
photons reduced to 50\%. Similarly, a nominal $\epsilon>100\%$ could correspond
to a "top heavy" IMF biased towards massive stars relative to Scalo, so that
the number of UV photons produced per collapsed baryons is increased.  We do
not attempt to summarize all the relevant uncertainties here (instead, see
Haiman \& Loeb 1997), and view $t_{\rm on}$ and $\epsilon$ as representative
parameters in the simplest possible model.  It is important to note, however,
that a combination of these parameters is constrained by the reionization
redshift $6\lsim z\lsim 40$ (Haiman \& Knox 1999), requiring that a few
ionizing photons per hydrogen atom be produced within this redshift interval.

Within this model, we start from a high redshift ($z=200$), and take small
redshift steps towards z=0.  At each step, the background flux is obtained by
summing the flux of all sources that are within the imaginary redshift screen
at $z_{\rm max}$ as discussed in \S~2.1, and satisfy the star formation
criterion of \S~3.2,

\begin{equation}
J_{21}(z,\nu)=\int_z^{z_{\rm max}} dz^\prime c \frac{dt}{dz^\prime}
j_{\nu^\prime}(z^\prime),
\label{eq:J21}
\end{equation}
where $j_{\nu}(z)$ is the total emissivity from all luminous
halos at redshift $z$ and frequency $\nu$,
\begin{equation}
j_{\nu}(z)=\frac{3.7\epsilon}{t_{\rm on}}\int_z^{z+dz(t_{\rm on})}
dz^\prime \frac{dF(z^\prime,T_{\rm crit}[z^\prime])}{dz^\prime},
\end{equation}
where $F(z,T_{\rm crit}[z])$ is the total fraction of baryons in
collapsed halos with virial temperatures above $T_{\rm crit}$.  
Here $t_{\rm on}$ is in units of seconds, and $j_{\nu}(z)$ is
in $10^{-21}~~{\rm erg~s^{-1}~cm^{-3}~Hz^{-1}~sr^{-1}}$.

Once this flux is obtained, we find the critical virial temperature numerically
for the new flux $J_{21}$ and redshift, using the information contained in
Figure~\ref{fig:Jmin}.

\section{Results}

\begin{figure*}[t]
\plotone{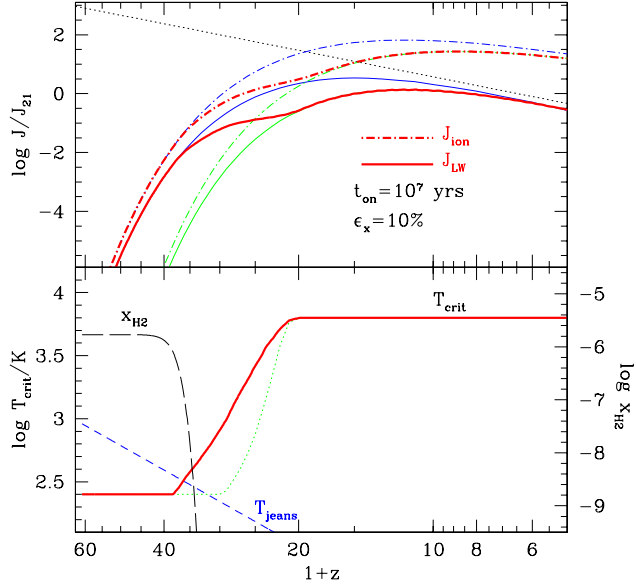}
\caption[Coupled evolution of $J$ and $T_{\rm min}$] {\label{fig:Res1}
  \footnotesize Coupled evolution of the background flux $J$, critical
  virial temperature $T_{\rm crit}$, and intergalactic ${\rm H_2}$
  fraction $x_{\rm H_2}$. We assume a UV production efficiency
  $\epsilon$=10\% and source life time $t_{\rm on}=10^7$yrs. {\it
    Upper panel}: The evolution of the ionizing flux at 13.6eV
  (dot--dashed lines) and the average flux in the LW bands (solid
  lines).  In both cases, the bottom and top curve assume a constant
  $T_{\rm crit}=10^{3.8}K$, and $10^{2.4}K$, respectively, and bracket
  our actual coupled solution, shown by the thick middle curve.  The
  diagonal dotted line shows the minimum flux for reionization,
  corresponding to one photon per hydrogen atom.  {\it Lower panel}:
  The evolution of the critical virial temperature in the coupled
  solution (thick solid curve), and using only the flux from halos
  with $T_{\rm vir}>10^{3.8}$K (dotted curve). The long--dashed line
  shows the evolution of the ${\rm H_2}$ fraction in the IGM.  The
  short--dashed line shows the virial temperature of the mass scale
  that is just Jeans unstable at each redshift.}
\end{figure*}

\begin{figure*}[t]
\plotone{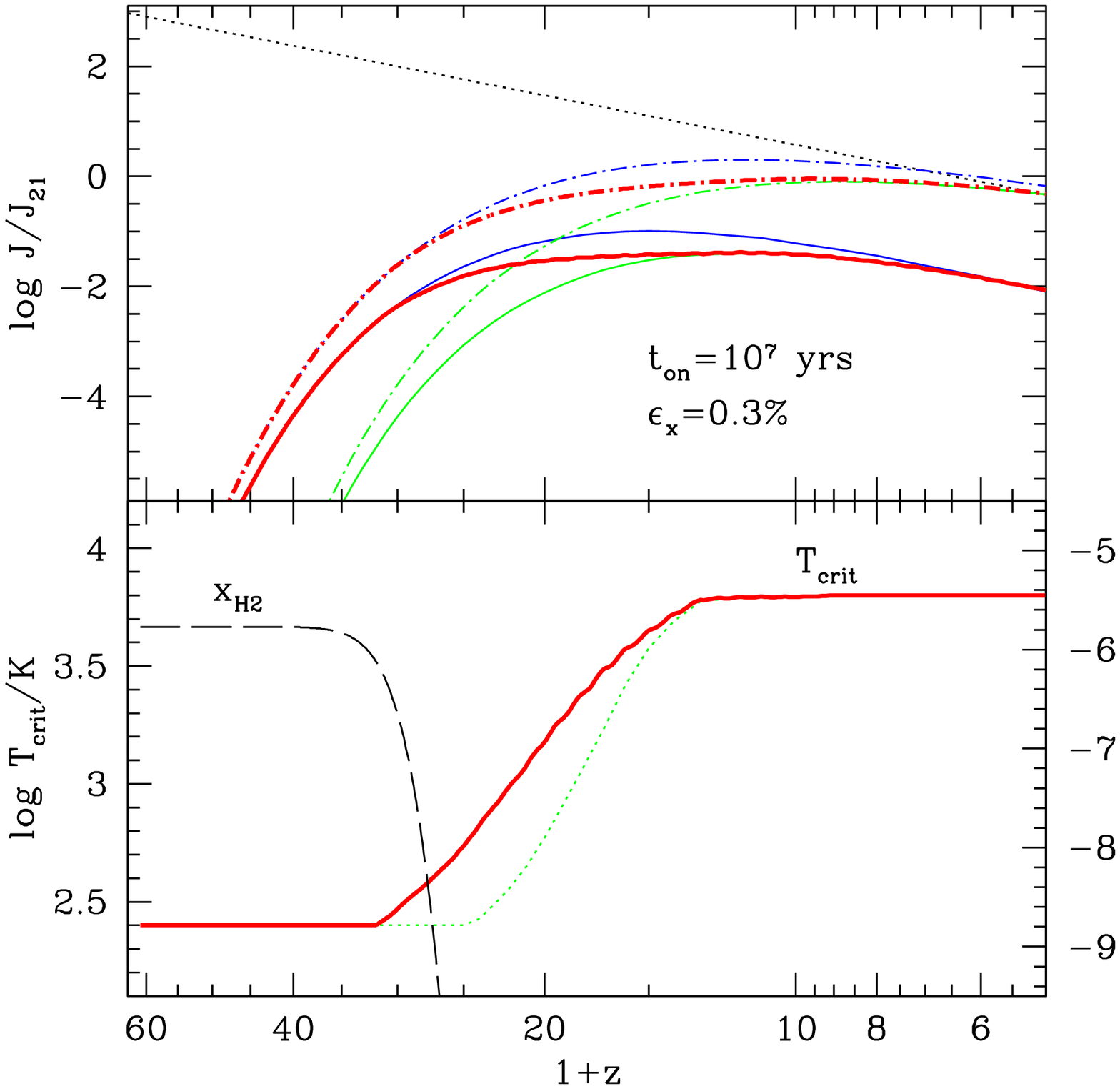}
\caption[Coupled evolution of $J$ and $T_{\rm min}$]
{\label{fig:Res2} \footnotesize Same as Figure~\ref{fig:Res1}, but
  assuming $\epsilon$=0.3\%.}
\end{figure*}

\begin{figure*}[t]
\plotone{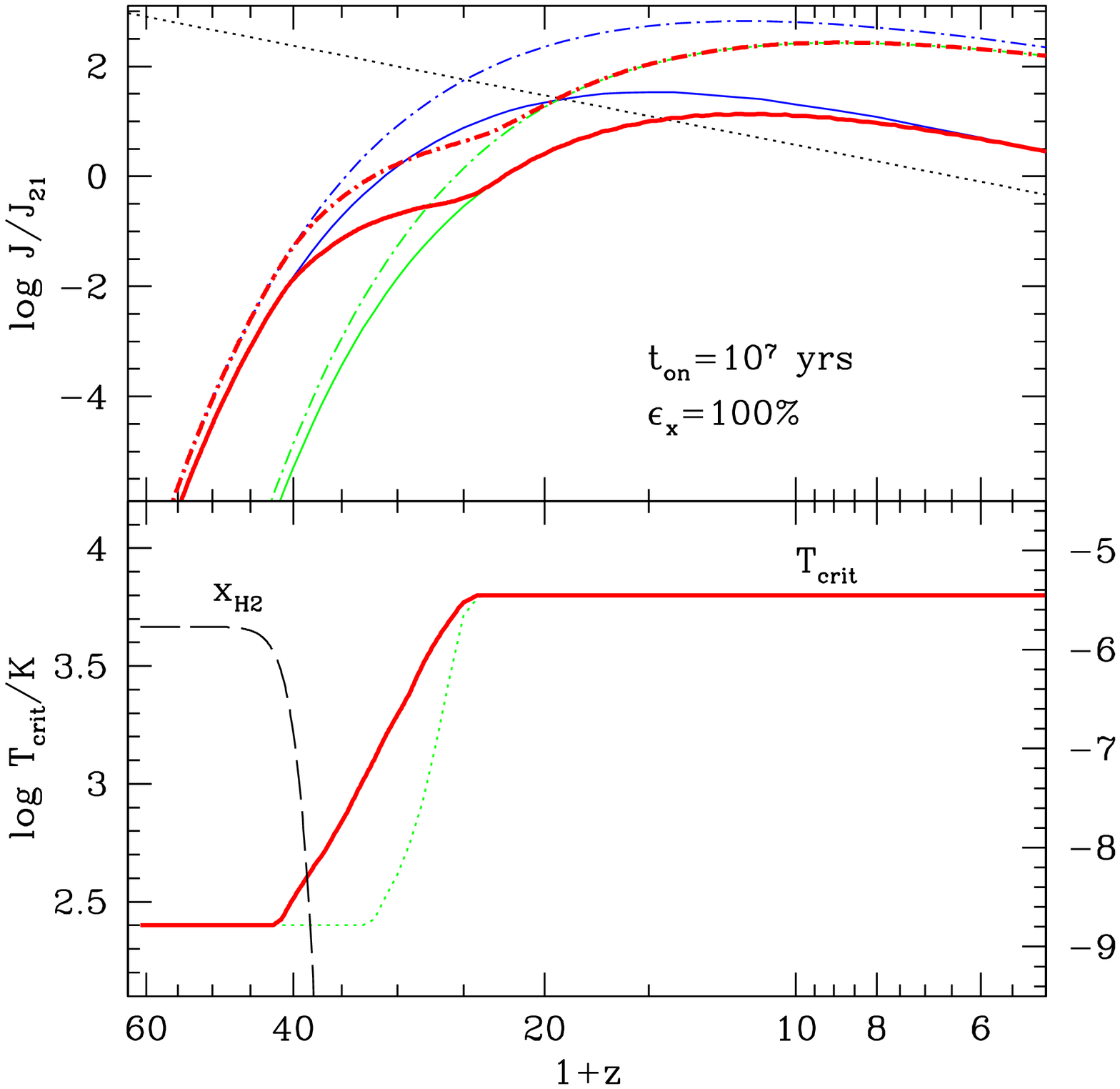}
\caption[Coupled evolution of $J$ and $T_{\rm min}$]
{\label{fig:Res3} \footnotesize Same as Figure~\ref{fig:Res1}, but
  assuming $\epsilon$=100\%.}
\end{figure*}

\begin{figure*}[t]
\plotone{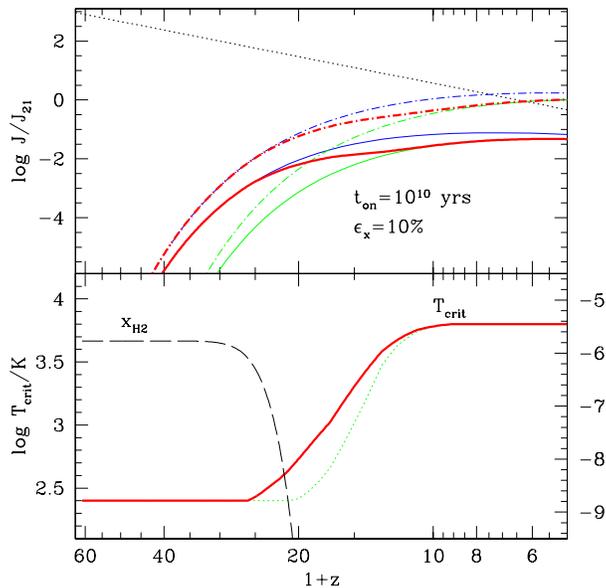}
\caption[Coupled evolution of $J$ and $T_{\rm min}$]
{\label{fig:Res4} \footnotesize Same as Figure~\ref{fig:Res1}, but
  assuming $t_{\rm on}=10^{10}$ yrs.}
\end{figure*}

\begin{figure*}[t]
\plotone{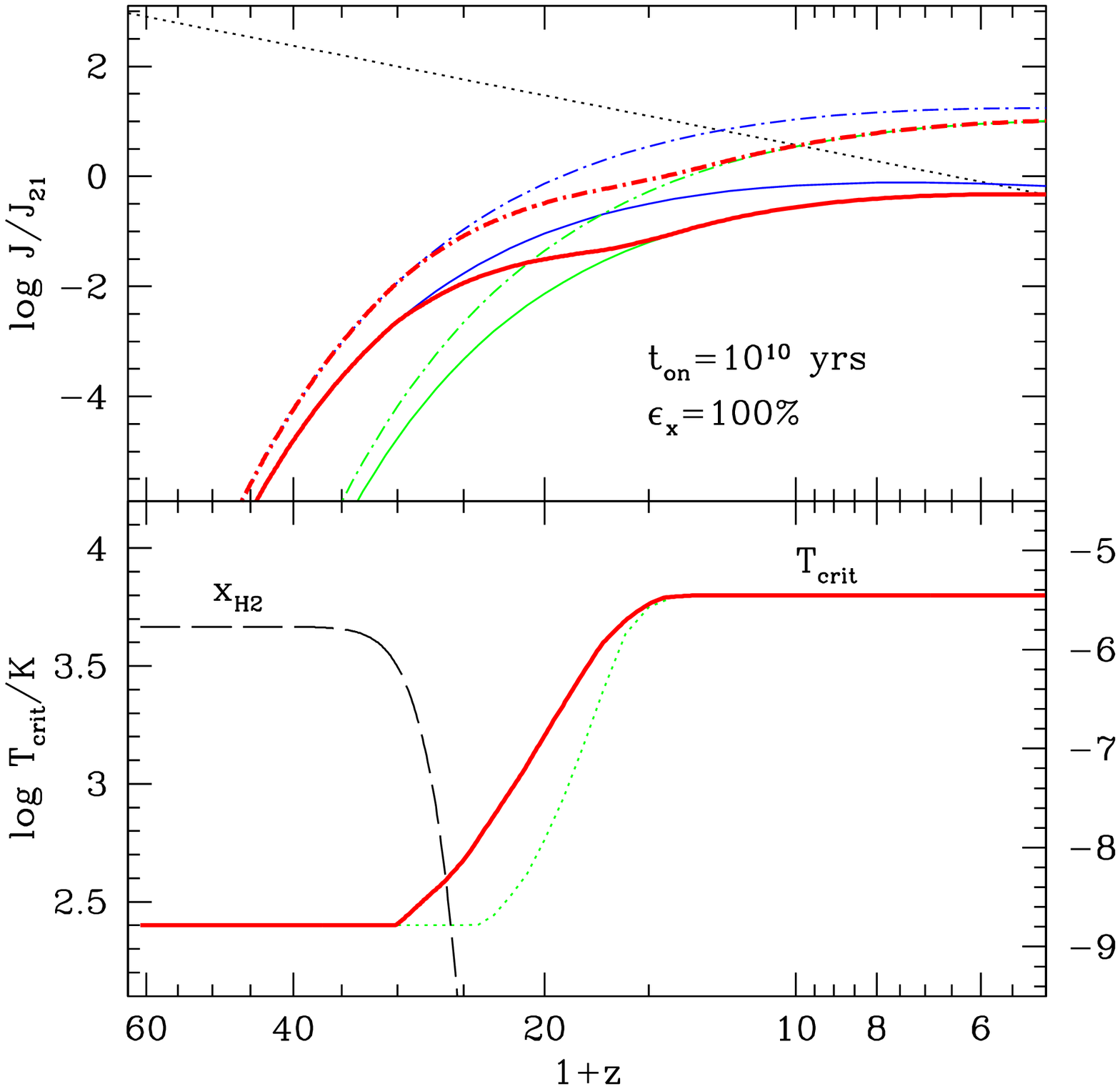}
\caption[Coupled evolution of $J$ and $T_{\rm min}$]
{\label{fig:Res5} \footnotesize Same as Figure~\ref{fig:Res4}, but
  assuming $\epsilon$=100\%.}
\end{figure*}

\begin{figure*}[t]
\plotone{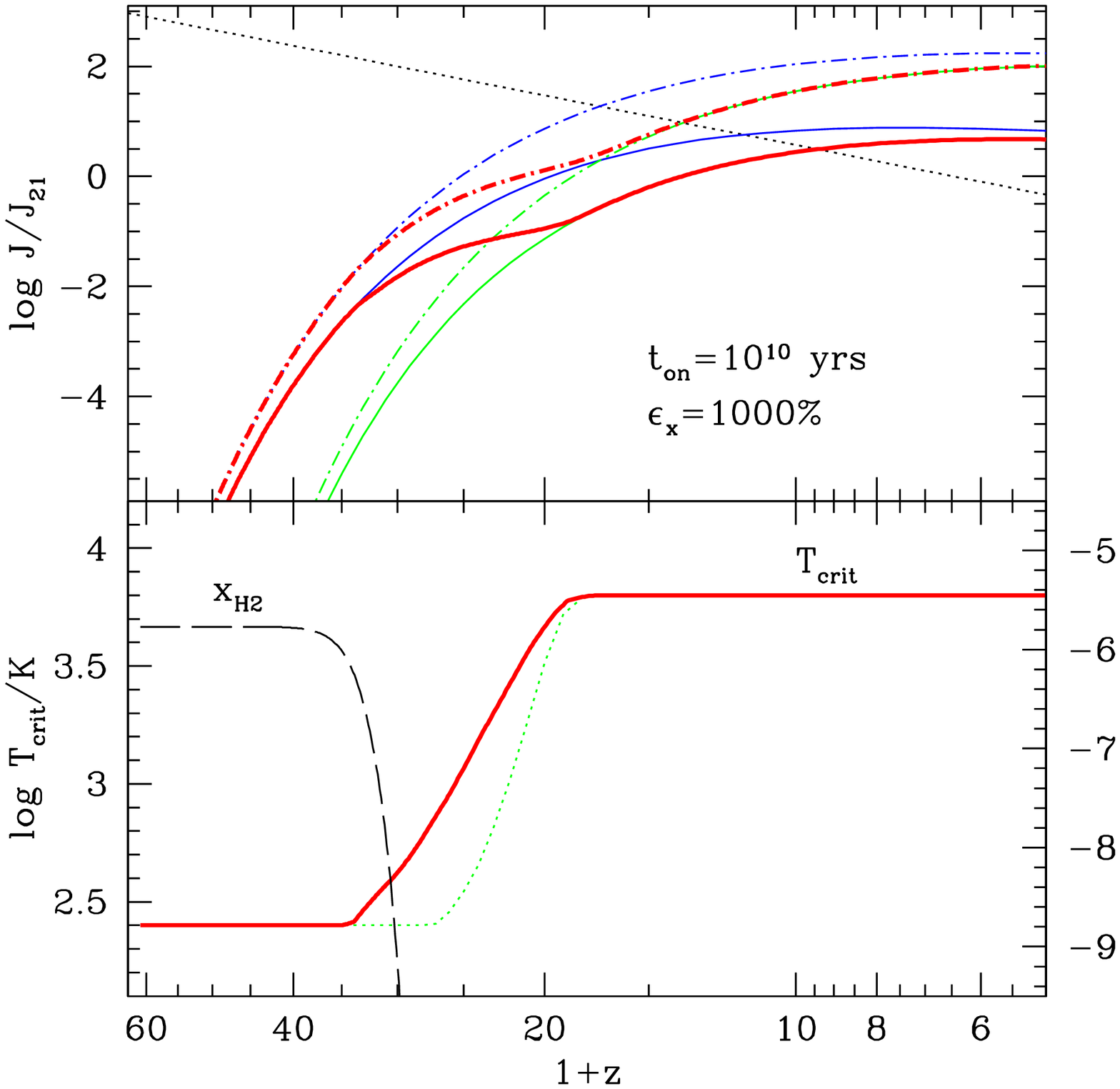}
\caption[Coupled evolution of $J$ and $T_{\rm min}$]
{\label{fig:Res6} \footnotesize Same as Figure~\ref{fig:Res4}, but
  assuming $\epsilon$=1000\%.}
\end{figure*}

Our main results are shown for six different combinations of lifetime and star
formation efficiency ($t_{\rm on}$,$\epsilon$) in
Figures~\ref{fig:Res1}--~\ref{fig:Res6}.  In each case, the figure shows the
evolution of the ionizing flux\footnote{Note that this flux is not actually
visible from a random vantage point until reionization is complete and all
sources can see each other at 13.6eV across the ionized IGM.} at 13.6 eV
($J_{\rm ion}$); the average flux in the LW bands, ($J_{rm LW}$); the critical
virial temperature $T_{\rm crit}$; and the intergalactic ${\rm H_2}$ fraction.

Figure~\ref{fig:Res1} shows the results in a model with ($t_{\rm
on}$,$\epsilon$) = ($10^7~{\rm yr}, 10\%$). The three solid curves in the top
panel show the evolution of $J_{\rm LW}$ in our model with feedback (middle
curve), assuming that the critical virial temperature is fixed at $T_{\rm
vir}=10^{2.4}$K (upper curve), and at $T_{\rm vir}=10^{3.8}$K (lower curve).
The solid line in the bottom panel shows the evolution of the actual critical
temperature, $T_{\rm crit}$, in the coupled solution. These curves demonstrate
that initially the small halos are allowed to form stars, there is no feedback,
and the flux builds up as in the constant $T_{\rm vir}=10^{2.4}$K case.  At
redshift $z\approx 35$, the feedback starts to be noticeable, as the flux
builds up to $J_{\rm LW}\approx 10^{-2}$, and $T_{\rm crit}$ rises above
$10^{2.4}$K.  By redshift $z=20$, $T_{\rm crit}$ reaches $10^{3.8}$K.  The
dot--dashed lines in the top panel show the corresponding evolution of the
ionizing flux, $J_{\rm ion}$; and the dotted line shows the minimum flux
required for reionization (corresponding to one photon per hydrogen atom).  As
the figure shows, reionization in this model occurs at $z\sim15$, and at this
redshift the contribution from the small halos to $J_{\rm ion}$ is only
$\sim$10\% percent\footnote{Our model shows that reionization occurs at
$z\lsim15$.  However, we assume an equality, because reionization is unlikely
to require more than a few ionizing photons per H atom (Miralda-Escud\'e et
al. 1999).}.  The long dashed line in the bottom panel shows the evolution of
the intergalactic ${\rm H_2}$ fraction.  As expected, $x_{\rm H_2}$ drops
rapidly to a negligible value by $z=35$, i.e.  before the ${\rm H_2}$--feedback
turns on.  This justifies our neglect of the modulation of the UVB by the
intergalactic ${\rm H_2}$.  The short--dashed line shows the virial temperature
of the mass scale that is just Jeans unstable at each redshift. This shows that
the initial IGM pressure plays a role only for the smallest halos at the
highest redshifts.

The next two figures illustrate the effect of changing the total star formation
efficiency (or alternatively, the efficiency of UV photon production per
collapsed baryon).  Figure~\ref{fig:Res2} is analogous to
Figure~\ref{fig:Res1}, but $\epsilon$ is lowered to $0.3\%$, the smallest
possible value that still produces reionization by $z=5$.  The evolution is
qualitatively similar to the $\epsilon=10\%$ case: the ${\rm H_2}$ feedback is
active during the redshift interval $13\lsim z \lsim 30$, before reionization
takes place at $z\approx 5$.  The star formation in small halos is suppressed,
and their contribution to the background flux at $z=5$ is only $\sim4\%$, less
than in the $\epsilon=10\%$ case.  This, however, is mostly an intrinsic
effect, rather than a result of the negative feedback: as shown by the
convergence of the top and bottom dashed curves, by the late time when
reionization occurs, the background ionizing flux is anyway dominated by the
large halos.  The figure also shows that the negative feedback has a smaller
effect than in the $\epsilon=10\%$ case (the middle and top dashed lines differ
by a smaller factor).  This is because the background flux builds up more
slowly, and the feedback is delayed to a later redshift. By this later
redshift, a larger fraction of the small halos have collapsed and formed stars.

Figure~\ref{fig:Res3} demonstrates the opposite effect of increasing the star
formation efficiency to a nominal $\epsilon=100\%$.  In this case, reionization
occurs earlier, at $z\approx 19$, and the ${\rm H_2}$ feedback operates at the
earlier epoch between $24\lsim z\lsim 42$.  As shown by the large difference
between the top and middle dashed curves, the negative feedback is more
pronounced, and it suppresses star formation in a larger fraction of the small
halos.  The total contribution of small halos to the ionizing flux at $z=19$ in
this case is $\sim 9\%$. In the absence of the feedback, the small halos would
dominate the flux by a factor of more than 10 at $z=19$, and reionization would
occur at $z\sim25$.

Figures~\ref{fig:Res4}--~\ref{fig:Res6} demonstrate the effect of increasing
the source lifetime, $t_{\rm on}$ to $10^{10}$ yrs.  This long lifetime would
also mimic the scenario in which constant mergers result in multiple recycling
of the gas (originally associated with a given halo) for star formation.
Making the nominal lifetime longer is equivalent to decreasing the star
formation rate, and has an effect similar to decreasing $\epsilon$: the buildup
of the background flux is now delayed (although not eventually suppressed as it
is when $\epsilon$ is decreased).  As a result, the reionization redshift is
delayed.  When $\epsilon$=10\%, then reionization is at $z\approx5$, and the
${\rm H_2}$ feedback takes place in the interval $10\lsim z \lsim 25$.  The
total contribution to the reionizing flux from small halos is nevertheless
still negligible, only about 2\%.  Figures~\ref{fig:Res5} and~\ref{fig:Res6}
show what happens when $\epsilon$ is increased to 100 or 1000\% (note that it
can not be decreased below 10\% because then reionization would occur at
$z<5$).  In both cases, the ${\rm H_2}$ feedback turns on before reionization,
and small halos contribute a negligible amount to the reionizing flux ($5\%$ at
$z=10$ and $7\%$ at $z=14$, respectively).

In summary, Figures~\ref{fig:Res1}--~\ref{fig:Res6} demonstrate that in all
cases, the ${\rm H_2}$ feedback suppresses star formation inside the small
halos, and that the contribution of the small halos to the reionizing flux is
always below 10\%.  If reionization occurs late, then the ${\rm H_2}$ feedback
makes little difference to the overall ionizing flux: the contribution from
small halos to the reionizing flux is then small in any case, since by this
late time, the collapsed baryon fraction, as well as the UV background, is
dominated by the large halos even in the absence of the ${\rm H_2}$
feedback. This is the case in Figures~\ref{fig:Res2} and~\ref{fig:Res4}.  This
would be especially true in different cosmologies, where structure forms
earlier, so that by the time of a late reionization, the dominance of larger
halos is even more pronounced.  Although not important for reionization in
these late--reionization models, the ${\rm H_2}$ feedback is still expected to
occur.  On the other hand, in models which would predict earlier reionization,
primarily due to the small halos, the negative ${\rm H_2}$ feedback is more
pronounced.  In these models, it is the feedback that limits the contribution
of the small halos to reionization, and delays reionization until larger halos
form.

Although we have assumed a universal star formation efficiency, stars in small
and large halos are likely to form differently, since the cooling mechanisms
are different in the two cases.  Thus, there is no a--priori reason to expect
the same star formation efficiency in these two different environments.  There
is a preliminary indication from 3--D simulations that the efficiency in small
halos might indeed be small ($<1\%$, Abel et al).  If the star formation
efficiency in the small halos were intrinsically much smaller than in the large
halos, then the contribution to the flux from the small halos would be small,
regardless of the ${\rm H_2}$--feedback.  We emphasize that the ${\rm H_2}$
feedback described here would still occur prior to reionization, since the flux
from the rare large halos alone is sufficient to suppress star formation inside
the small halos.  Note that this scenario is different from the one presented
in, e.g.  Figure~\ref{fig:Res1}, since in that case the small halos exert a
feedback on their own population, while here the ``feedback'' is from a
distinct population of larger halos. This is demonstrated explicitly in
Figures~\ref{fig:Res1}--~\ref{fig:Res6}, where the dotted curves in each bottom
panel shows the critical virial temperature $T_{\rm crit}$, obtained from the
UV background due only to the large halos.  As these figures show, in all
cases, $T_{\rm crit}$ still rises above $10^{2.4}$K and reaches $10^{3.8}$K
before reionization.

It is not surprising that the ${\rm H_2}$ feedback generically occurs in any
model that satisfies the reionization constraint by $z=5$. This is because, as
we have shown in \S~3.3, the suppression of the ${\rm H_2}$ abundance and star
formation inside clouds with the relevant range of densities and temperatures
requires a UV flux that is only a fraction $10^{-2}-10^{-4}$ of the minimum
flux level required for reionization (one photon per hydrogen atom).  As the
UVB builds up to the reionizing level, it must necessarily first reach the
level at which the ${\rm H_2}$ feedback sets in.  

\section{Early Mini--Quasars and the Effects of X--rays}

\begin{figure*}[t]
\plotone{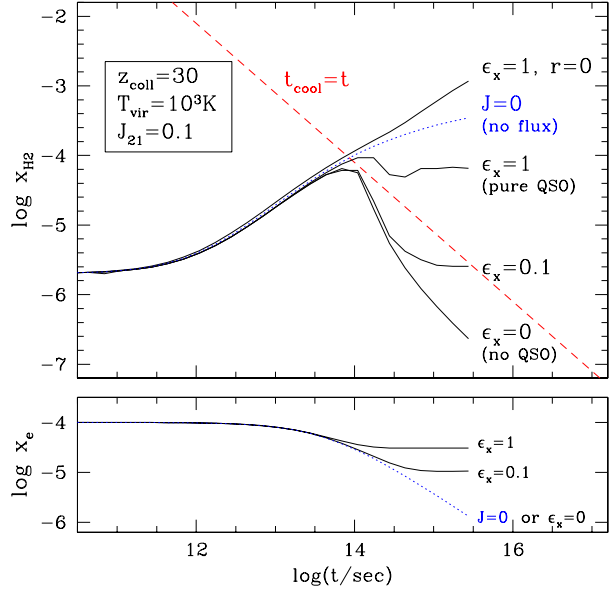}
\caption[Effect of X--rays] {\label{fig:xrays} \footnotesize Effect of
  X--rays on the evolution of the ${\rm H_2}$ abundance.  The top
  panel is similar to Figure~\ref{fig:evol}, except that $z_{\rm
    coll}=30$, and $J_{21}=0.1$, and we have added an X--ray
  contribution to the background flux.  The curve labeled J=0 shows
  the evolution of the ${\rm H_2}$ abundance in the absence of any
  flux.  The label $\epsilon_{\rm x}$ on the other curves is a measure
  of the X--ray/UV flux ratio, where $\epsilon_{\rm x}=0$ corresponds
  to a pure stellar spectrum with no X--rays, and $\epsilon_{\rm x}=1$
  corresponds to a pure ``mini--quasar'' spectrum with a spectral
  index of -1. The bottom panel shows the evolution of the free
  electron fraction in each case.}
\end{figure*}

\begin{figure*}[t]
\plotone{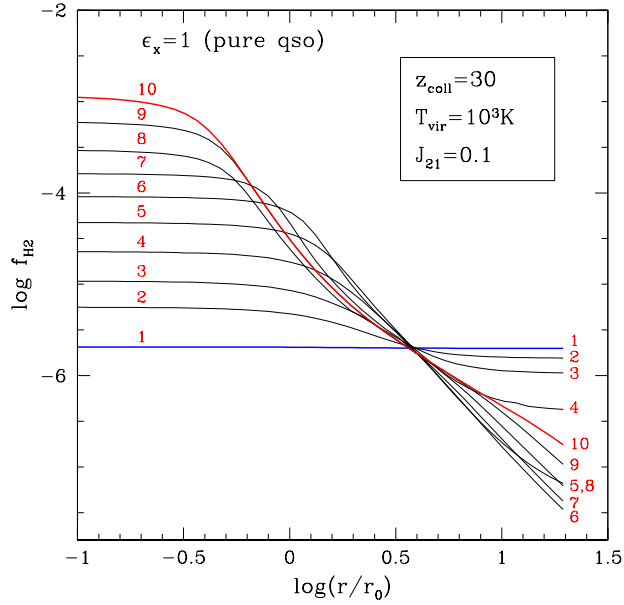}
\caption[Effect of X--rays] {\label{fig:xrayprof} \footnotesize Effect
  of X--rays on the profile of the ${\rm H_2}$ abundance, for the same
  truncated isothermal sphere as in Figure~\ref{fig:xrays}. Increasing
  numbers correspond to equal time--steps in logarithmic time, upto a
  Hubble time of $\sim 3\times10^{15}$ sec. Note that the ${\rm H_2}$
  fraction still rises within the core $r<r_0$ by an order of
  magnitude.}
\end{figure*}

So far, we have assumed that the early sources have ``stellar'' spectra, and
have simply truncated the observed flux above 13.6 eV.  Since all of our
results above depend on the flux only in a narrow frequency range below 13.6eV,
they are indeed insensitive to the precise shape of the spectrum.

An alternative possibility is that a halo collapsing at high redshift forms a
central black hole, and turns into a "mini--quasar", with a hard spectrum
extending into the X-ray regime (Haiman \& Loeb 1998).  The ionization
cross--section of neutral hydrogen drops rapidly with frequency (as
$\nu^{-3}$), and at frequencies above $h\nu\gsim 1$keV, the neutral IGM is
optically thin.  If the spectra of the early sources have a non--negligible
contribution at these frequencies, they would establish an early X--ray
background (XRB), and change the chemistry qualitatively.

The importance of the X--rays is that they provide additional free electrons,
both by direct photo--ionization of H and He, as well as indirectly, by
collisional ionization of H by the fast photo--electrons.  The resulting extra
free electrons catalyze the formation of $H^-$, and therefore tend to increase
the ${\rm H_2}$ abundance.  In an earlier paper (HRL96), we have found that
inside dense regions ($n_{\rm H} \gsim 1~{\rm cm^{-3}}$), the overall effect of
a background flux extending to hard X--rays is to enhance, rather than to
suppress, the ${\rm H_2}$ abundance.  The central densities of centrally
condensed virialized clouds are expected to reach $n_{\rm H} \sim 1~{\rm
cm^{-3}}$.  In the case of the TIS solutions adopted here, this density is
exceeded for halos that collapse at $z_{\rm coll}\gsim 7$, and therefore a
positive feedback is expected when the UV and X--ray backgrounds together
illuminate these halos.

In order to illustrate the effect of the X--rays, we have recomputed the
evolution of the ${\rm H_2}$ profiles in the TIS models.  These computations
are identical to the ones discusses in \S~3.1, except that the flux in
equation~\ref{eq:flux} is modified to
\begin{equation}
\small{
J_\nu =
\left\{\matrix{
J_{21}\left(\frac{\nu}{\nu_{\rm H}}\right)^{-1}
\times {\rm sawtooth},
 \hfill& h\nu<13.6 {\rm eV}  \hfill\cr
\epsilon_{\rm x}J_{21}\left(\frac{\nu}{\nu_{\rm H}}\right)^{-1}
e^{(-10^{22}[\sigma_{\rm H,\nu}+0.08\sigma_{\rm He,\nu}])},
 \hfill& 13.6 - 10 {\rm keV}  \hfill\cr
0,\hfill& h\nu>10 {\rm keV}  \hfill\cr} \right.}
\label{eq:XRB}
\end{equation}
Here $\sigma_{\rm H,\nu}$ and $\sigma_{\rm He,\nu}$ are the ionization
cross--sections of H and He, and the exponential factor mimics the absorption
above 13.6 eV by the neutral IGM by assuming a fixed hydrogen column density of
$N_{\rm H}=10^{22}~{\rm cm^{-2}}$ and a helium column density a factor of 0.08
smaller.  The fiducial value for $N_{\rm H}$ was chosen based on a comparison
of background fluxes computed in our Press--Schechter model with and without
absorption by neutral H and He at $z=20$.  The units of $J_{21}$ are
$10^{-21}~~{\rm erg~s^{-1}~cm^{-2}~Hz^{-1}~sr^{-1}}$.  We have also introduced
a parameter $\epsilon_{\rm x}$ to characterize the ratio of X--ray to UV flux:
for pure stellar sources, $\epsilon_{\rm x}=0$; for typical quasar spectra
(Elvis et al.), $\epsilon_{\rm x}\sim 1$.  Values of $\epsilon_{\rm x}$
in--between 0 and 1 could describe either miniquasars with softer spectra, or a
mixed population of stars and mini--quasars.  We have verified that our
conclusions are not sensitive to the upper energy cutoff, provided this cutoff
is above $\sim$1keV.

In Figure~\ref{fig:xrays}, analogously to Figure~\ref{fig:evol}, we show the
evolution of the ${\rm H_2}$ fraction at the core radius of a TIS. For this
illustrative example, we assumed that the TIS has $z_{\rm coll}=30$ and $T_{\rm
vir}=10^3$K, and that it is illuminated by a flux with $J_{21}=0.1$ and
$\epsilon_{\rm x}=0,0.1,$ and 1, corresponding to increasing amounts of X--ray
contribution.  As in the example shown in Figure~\ref{fig:evol}, in the absence
of any flux the ${\rm H_2}$ fraction rises continuously, in this case to
$\sim10^{-3.5}$ (dotted curve).  If the UVB is turned on, the final ${\rm H_2}$
fraction is suppressed to $\sim10^{-7}$ (bottom solid curve).  However, if
increasing amounts of X--rays are also added (solid curves, second and third
from the bottom), then the ${\rm H_2}$--suppression (relative to the dotted
curve) is less pronounced.

We find that a $\sim$ 10\% X-ray contribution ($\epsilon_{\rm x} \approx 0.1$)
is enough to keep the ${\rm H_2}$ fraction at a high enough level
($\sim10^{-5.5}$) that satisfies our star formation criterion. This is seen in
Figure~\ref{fig:xrays}, as the solid curve labeled $\epsilon_{\rm x}=0.1$
reaches the short--dashed curve.  For still larger X--ray contributions, the
final ${\rm H_2}$ fraction is even higher.  It is interesting to note that
deeper inside the cloud, the sign of the feedback is entirely reversed, and the
${\rm H_2}$ abundance is {\it increased} relative to the case with no flux.
The top solid curve in Figure~\ref{fig:xrays} shows the ${\rm H_2}$ fraction at
the center $r=0$, rather than at the core radius, in the $\epsilon_{\rm x}=1$
case.  Indeed, $x_{\rm H_2}$ rises above $10^{-3}$, and reaches a $\sim3$ value
than it would in the absence of any background flux.  Figure~\ref{fig:xrayprof}
shows the profile of $x_{\rm H_2}$ across the cloud at various time--steps, and
demonstrates that the ${\rm H_2}$ fraction still rises within the core
($r<r_0$) by over an order of magnitude. As described above, the enhancement of
the final ${\rm H_2}$ fraction is due to the additional free electrons produced
by the X--rays.  The bottom panel in Figure~\ref{fig:xrays} shows the evolution
of the electron fraction in each case: as expected, $x_{\rm e}$ is increased by
the presence of X--rays.

By experimenting with different combinations of $z_{\rm coll}$ and $T_{\rm
vir}$, we have found that the situation shown in Figure~\ref{fig:xrays} is
typical of collapsed clouds with $5<z_{\rm coll}<50$ and $10^{2.4}{\rm K}<
T_{\rm vir}<10^4$K.  Generally, a $\sim 10\%$ contribution by X--rays to the
flux is sufficient to keep the ${\rm H_2}$ fraction high enough to allow star
formation, and hence to eliminate the negative feedback caused by the UV
background.  To arrive at this conclusion, we have imposed our star--formation
criterion at the core radius $r=r_0$. However, we have also found that at still
smaller radii, the ${\rm H_2}$ fraction is typically much larger, and can rise
above the value expected in the absence of any background flux. Inside the
core, net feedback on the ${\rm H_2}$ fraction is therefore positive, rather
than negative.  In summary, we expect that if the early background had a
non--negligible contribution from ``mini--quasars'' with hard spectra,
extending to $\sim 1$ keV, then the negative ${\rm H_2}$ feedback described
here would not occur.

\section{Conclusions}

We have examined the build--up of the UV background in hierarchical models,
based on the Press--Schechter formalism, and its effect on star formation
inside small halos that collapse prior to reionization.  We have confirmed our
previous conjecture that there exists a negative feedback before reionization.
Star formation inside small halos shuts off prior to reionization, as an early
UV background below builds up below 13.6 eV, and suppresses the ${\rm H_2}$
abundance in these halos.

We have also found that as a result, the contribution to the reionizing flux
from the small halos is limited to be less than a $\sim$10\% percent in any
hierarchical model.  It is possible that the star formation efficiency via
${\rm H_2}$ cooling in small halos is intrinsically small, as indicated by
recent 3--D simulations (Abel et al.).  The efficiency can also be kept low by
internal feedback mechanisms inside each halo, such as destruction of ${\rm
H_2}$ molecules by an internal UV source (Silk 1977, Omukai \& Nishi 1999), or
blow--out of the gas by supernovae (Mac Low \& Ferrara 1998).  In this case,
the contribution of the small halos to reionization can be negligible
regardless of the existence of the negative feedback.  However, we have shown
that the negative feedback would still occur from the flux of the large halos
alone.

Another possibility is that even though the integrated star formation
efficiency is high, the star formation rate is low, so that the UV flux is
built up with a significant delay relative to the collapse of the halos. In
this scenario, reionization occurs late (near $z\sim 5$), when the collapsed
baryon fraction is already dominated by large halos. Again, in this case, the
negative feedback would still occur early on, but would not have a significant
effect on reionization.

More generally, we have shown that the contribution of the small halos to the
reionizing flux is small in essentially any model where the first collapsed
halos do not produce hard (E$\gsim$1keV) X--rays.  The reason for this is
that the flux necessary to suppress the ${\rm H_2}$ abundance and cooling in
small halos is $10^{2-4}$ times lower than the minimum flux necessary for
reionization, corresponding to an one UV photon per baryon.

Finally, we have shown that the negative feedback would be turned into a
positive one if there was a significant ($\gsim$10\%) contribution to the early
UV background from ``mini--quasars'' with hard spectra extending to the X--rays
(upto $\sim$1keV).  In this case, the X--rays catalyze the formation of
${\rm H_2}$ molecules by providing additional electrons, and the overall effect
is to enhance, rather then to suppress, the ${\rm H_2}$ abundance.  The
evolution of the small early halos is therefore distinctly different in stellar
and mini--quasar models: this may help shed light on the question of what type
of sources were responsible for ending the ``dark age'', and reionized the
Universe.

\acknowledgements

We thank Jordi Miralda-Escud\'e and Mike Fall for stimulating discussions.
This research was supported by the DOE and the NASA grant NAG 5-7092 at
Fermilab. Tom Abel acknowledges support from NASA grant NAG5-3923.


\appendix
\section{The Relevant Lyman--Werner Lines of ${\rm\bf\protect H_2}$ Molecules}

\begin{figure*}[t]
\epsscale{0.5}
\plotone{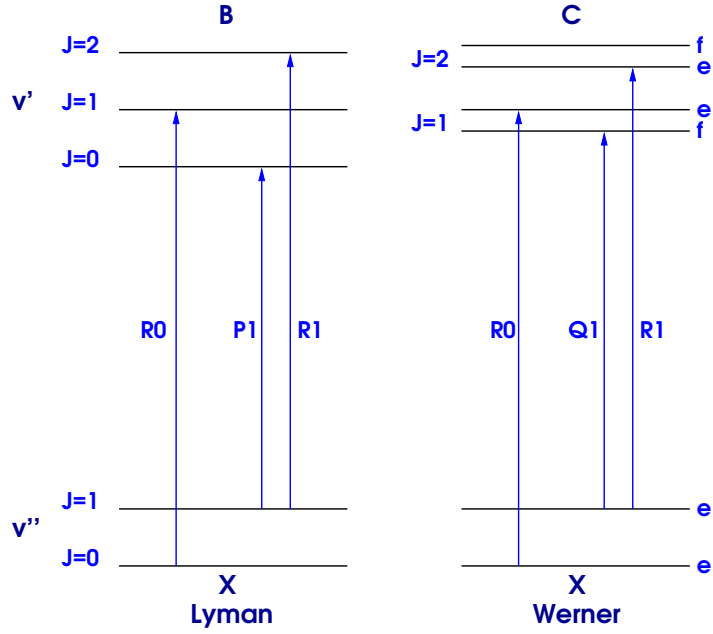}
\caption{\label{fig:term}{\footnotesize  Illustration of all possible
    absorption lines from the lowest energy states of ortho and para
    \HH.  }}
\end{figure*}

The bulk of the intergalactic H$_2$ is in the lowest states vibrational and
rotational state of the ground electronic state. Hence only v=0 with J=0 (para)
and 1 (ortho) are populated. There are numerous Lyman Werner band transitions
that can yield photo-dissociation via a decay from excited electronic states to
the continuum of two distinct H atoms.  These are typically classified as R or
P transitions, depending on whether the rotational quantum number changes by -1
or +1 from the electronic ground to the electronic excited state (angular
momentum must change due to parity conservation). Hence, assuming that both
para and ortho hydrogen molecules are abundant in their ground states, all R(0)
[i.e.  $J=0 \rightarrow J=1$], R(1) [i.e.  $J=1\rightarrow J=2$], and P(1)
[i.e.  $J=1 \rightarrow J =0$] transitions are possible, for each excited
vibrational level. The above is true both for the Lyman and Werner bands. For
the Werner band ($X\, ^1\Sigma_g^+ - C\, ^1\Pi_u$), there is an additional
Q--branch in which $\triangle J=0$ is allowed\footnote{this is due to the
different orbital angular momentum along the inter--nuclear axis of the two
states, i.e. $\Sigma - \Pi$ for the Werner but $\Sigma - \Sigma$ for the Lyman
transition ($\Lambda$ doubling)}.  Figure~\ref{fig:term} illustrates all
possible transitions from the lowest energy states of ortho and para \HH.

\begin{figure*}[t]
\epsscale{0.6}
\plotone{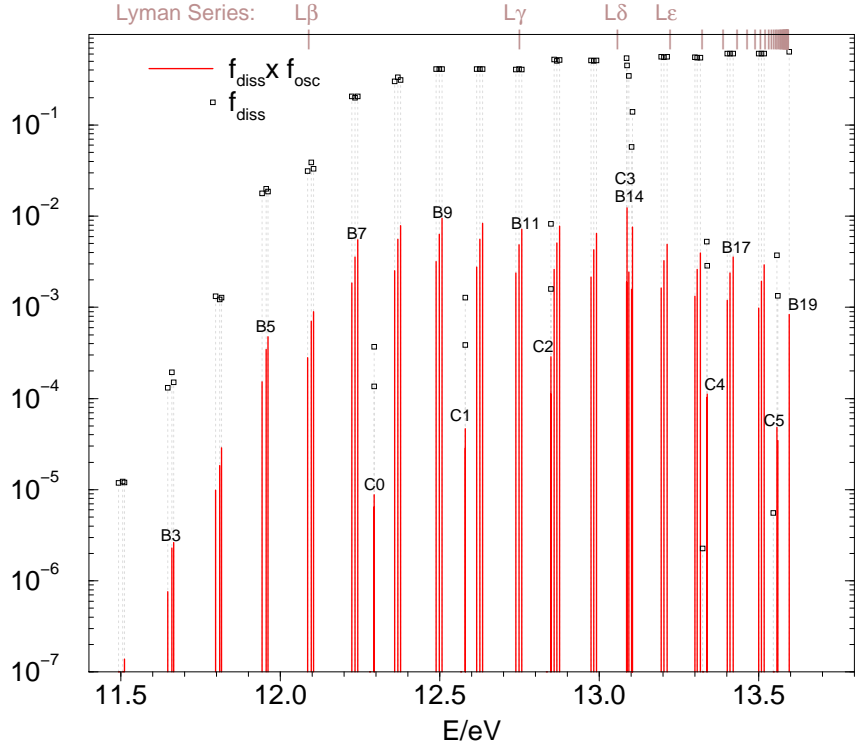}
\caption{\label{fig:LW}{\footnotesize The relevant Lyman Werner bands
    for intergalactic \HH\ photo-dissociation. The solid columns
    indicate the value of $f_{diss}\times f_{osc}$ and the dotted ones
    show the dissociation fractions. Lines of the Lyman (Werner)
    system are labeled by B$v'$ (C$v'$) where $v'$ is the vibrational
    quantum number of the excited state. The location of the atomic
    hydrogen Lyman series are indicated on the top of the graph.  }}
\end{figure*}

There is a total of 76 possible transitions from the lowest roto-vibrational
state of ortho and para molecular hydrogen that are below the Lyman limit of
atomic hydrogen (13.6eV).  Figure~\ref{fig:LW} gives an overview of these 76
lines that are relevant for cosmological \HH\ photo--dissociation.  The
quantity $f_{osc}\times f_{diss}$ quantifies the contribution of an individual
line to the overall photo--dissociation.  The $J$--dependent dissociation
fractions in this figure were taken from Abgrall~\etal(1992). From their work
one sees that the dissociation fractions of the Werner Q--branch (which couples
to the $\Pi^-$ state) are extremely small. For the R--branch, however, the
$f_{diss}$ are quite significant (especially for $v'=3$ of the Werner band).
As the figure shows, photons with $h\nu \lsim 11.9\eV$ do not contribute
significantly to the dissociation rate due to their small $f_{osc}$.  In
addition, the Werner bands with $v'\neq3$ do not contribute much to the
dissociation rate.

\vfill
\eject
\end{document}